\documentclass[prd,aps,twocolumn,superscriptaddress]{revtex4}
\usepackage{amsmath, amsthm, amssymb,float}
\usepackage{amsfonts}
\usepackage{graphicx}
\usepackage{dcolumn}
\usepackage{bm}
\usepackage{textcomp}
\usepackage[normalem]{ulem}

\usepackage[titletoc,title]{appendix}

\usepackage{color}
\begin{document}

\title{Highly Radiating Charged Particles in a Strong Electromagnetic Field}

\author{P. Sasorov} \email{pavel.sasorov@eli-beams.eu}
\affiliation{Institute of Physics ASCR, v.v.i. (FZU), ELI-Beamlines Project, 182 21 Prague, Czech Republic}
\author{M. Jirka} 
\affiliation{Institute of Physics ASCR, v.v.i. (FZU), ELI-Beamlines Project, 182 21 Prague, Czech Republic}
\affiliation{Faculty of Nuclear Sciences and Physical Engineering, Czech Technical University in Prague, Brehova 7, 115 19 Prague, Czech Republic}
\author{S.V. Bulanov} \email{sergei.bulanov@eli-beams.eu}
\affiliation{Institute of Physics ASCR, v.v.i. (FZU), ELI-Beamlines Project, 182 21 Prague, Czech Republic}

\begin{abstract}
We consider highly radiating ultra-relativistic electrons in a strong external electromagnetic field. High intensity radiative losses and consequent $e^+e^-$-pair production, appearing in the frame of quantum electrodynamics, determine indirectly an effective mass of electrons in the strong field. We calculate a leading term of the mass at asymptotically high energies of electrons in a strong constant field. We do not use any perturbation theory based on a low ratio of the effective mass correction to the electron mass $m_e$, but only the original small parameter of quantum electrodynamics,  the fine structure constant, $\alpha$.  The analogous
result is obtained for an effective mass of photons. These results resolve at least partially a very long-lasted controversy originating from traditional and straightforward application of perturbative approaches for description of highly radiating ultra-relativistic charged particles in a very strong electromagnetic fields.
\end{abstract}

\maketitle

\nopagebreak
\section{Introduction}

Cascading creation of $\gamma$-quanta and electron-positron pairs by highly energetic electrons (positrons) and $\gamma$-quanta in a  strong electromagnetic field in a vacuum attracts  a lot of attention~\cite{BK08,Pi12,Gon21} (see also references therein). This process can lead to formation of ultra-relativistic plasma composed of $e^\pm$-pairs and $\gamma$-quanta. The cascades were described theoretically~\cite{Sok10,Pi12,St13,B15,Mag19,Gon21} and with computer simulations~\cite{Mag19,Sam21,Qu21,Gon21}. Possibilities of investigation of such processes in forthcoming experiments are considered in Refs.~\cite{Zh20,Gon21}.  High attention to these problems is caused at least partially by modern astrophysical problems related to the physics of pulsars~\cite{Mel16,Cer17}, magnetars~\cite{Ka17} and of possible sources of fast radio bursts~\cite{Co19}.

Regardless of a lot of interesting physical problems, mentioned above and concerning strongly radiating particles, we are interested here in a more fundamental problem relevant to such the particle dynamics in a strong electromagnetic field. There is a long-lasted fundamental problem of Quantum ElectroDynamics (QED) in a strong external electromagnetic field. Most results obtained in this area are obtained really with a special version of perturbation theory adjusted especially for investigation of QED effects in a strong external field. The perturbation theory, based on the Furry picture~\cite{Fu51}, is being developed for more than a half of century and is widely used. For the references see Refs.~\cite{R70,R72,N69,N79a,N79b,R79,F18,F17,Mi20,R86,Il19,PL20,Fe22} and literature cited therein. It is widely accepted that the previously formulated theory will not be applicable for ultra-relativistic particles with sufficiently high energies in a sufficiently high intensity electromagnetic field.

This limitation can be expressed as
\begin{equation}\label{E010}
\alpha \chi^{2/3} < 1\mbox{~~~and~~~}\alpha \kappa^{2/3} < 1\, .
\end{equation}
Here $\alpha=e^2$ is the fine structure constant, and the dimensionless relativistic invariants $\chi$ and $\kappa$, describing intensity of interaction of electrons and photons with the strong external electromagnetic field, are defined as
\begin{equation}\label{SF514}
\chi^2=\frac{e^2 F_{\nu\mu}F^{\mu\lambda}p^\nu p_\lambda}{m_e^6}
;\quad
\kappa^2=\frac{e^2 F_{\nu\mu}F^{\mu\lambda}k^{\nu}
k_\lambda}{m_e^6};
\end{equation}
where $p^\mu$ and $k^\mu$ are 4-momenta of electrons and photons, respectively, $F_{\mu\nu}$ is tensor of the  electromagnetic field, and $e$ is electric charge of electrons~\cite{un}.

There are two related to each other reasons of appearing the constraints~(\ref{E010}) on the applicability of the pertubative QED in a strong field. Calculations~\cite{R70,R72} of the mass operator for electrons give a correction to the electron mass of the order of $\alpha m_e \chi^{2/3}$ at $\chi\gg 1$. The correction becomes of the order of the electron mass at $\alpha\chi^{2/3}\sim 1$.  As a result, authors of these papers expressed reasonable concern that higher correction cannot be obtained with the perturbation theory. An analogous situation takes place for the polarization operator for photons~\cite{N69,R70}. Investigations~\cite{N79a,N79b,R79} of higher order terms in the perturbation theory series show that indeed a small parameter of the QED perturbation theory in a strong field is equal to $\alpha\chi^{2/3}$ or $\alpha\kappa^{2/3}$, starting from a certain order of the perturbation theory. These results mean that the perturbation theory for polarization and mass operators breaks really at $\alpha\chi^{2/3}$ and $\alpha\kappa^{2/3}\sim 1$. It is even more or less widely accepted~\cite{F18,F17,Il19,PL20,He21} that a real, physical peculiarity in QED in a strong external field could be possible at  $\alpha\chi^{2/3}$ and $\alpha\kappa^{2/3}\gtrsim 1$. However,  recent results~\cite{Mi20,Tor21,Mi21} indicate that resummation of a class of diagrams of the perturbation theory, giving formally divergent result at $\alpha\chi^{2/3}\sim1$ for the electron mass operator, may give a finite result at high $\alpha\chi^{2/3}$. These findings inspire us to explore a simpler way for extension of known results for the effective electron and photon masses to higher $\alpha\chi^{2/3}$ and $\alpha\kappa^{2/3}$.

In the present paper we consider a way for calculation of the mass and polarization operators at arbitrary $\alpha\chi^{2/3}$ and $\alpha\kappa^{2/3}$, avoiding as much as possible argumentations based on the perturbation approaches. We consider different processes in a constant electromagnetic field, when $F_{\mu\nu}$ does not depend on space-time, with both Poincar\'e invariants vanishing: $F_{\mu\nu}F^{\mu\nu}=\varepsilon_{\mu\nu\lambda\rho}F^{\mu\nu}F^{\lambda\rho}=0$.
Here, $\varepsilon_{\mu\nu\lambda\rho}$ is the Levi-Civita symbol in four dimensions with $\varepsilon_{0123}=1$. Such kind of the electromagnetic field is called as a `crossed' field and approximates locally a plane electromagnetic wave with a very low frequency. This approach is called also as `constant field approximation'.

\section{Radiation correction to electron mass at $\chi\gg1$}

We consider for convenience an ultra-relativistic electron in the crossed field with $E\ll E_s =m_e^2/e$. Here $E$ is absolute value of the electric field of the crossed field. We restrict ourselves only to the case $\chi\gg 1$. 

Differential probability to emit one photon per one electron and per unit time averaged over spin orientation of the initial electron is calculated in~\cite{NR64,NR67,R70,R79}. It is equal to
$$
dP=\alpha\frac{m_e^2}{p^0}\left(\frac{u}{2\chi}\right)^{1/3}\frac{du\, d\tau}{(1+u)^3}\,
$$
$$
\times\biggl\{\left[u^2+\tau^2(u^2+2u+2)\right]\mbox{Ai}^2(y)
$$
\begin{equation}\label{SF510}
+\left(\frac{2\chi}{u}\right)^{2/3}(u^2+2u+2)\,
\mbox{Ai}^{\prime\,2}(y)\biggr\}\, ,
\end{equation}
where
\begin{equation}\label{SF512}
u=\frac{\kappa}{\chi^\prime};\quad
\tau=\frac{eF_{\lambda\mu}^*k^{\prime\, \lambda}p^\mu}{m_e^4\, \kappa}
;\quad
y=\left(\frac{u}{2\chi}\right)^{2/3}(1+\tau^2);
\end{equation}
$p^\lambda$, $p^{\prime\lambda}$ and $k^\lambda$ are 4-momenta of incoming and outgoing electrons and emitted photons respectively. Here $F^{\mu\nu}$ is covariant tensor of the constant electromagnetic field, and $F_{\mu\nu}^*=\varepsilon_{\mu\nu\lambda\rho}F^{\lambda\rho}/2$ is its dual version. $\chi^\prime$ differs from $\chi$ by the substitution $p\to p^\prime$ in the definition of $\chi$. Meanwhile, $\chi=\kappa+\chi^\prime$. Hence, $u$ describes redistribution of initial `energy' of incoming electron between outgoing photon and outgoing electron.

When $\chi\gg1$, and we neglect relatively small contribution from the emitted photons with almost $\kappa\to \chi$ and $\chi^\prime\sim 1$, so that both $\chi$ and $\chi^\prime\gg1$, then the expression~(\ref{SF520}) for $dP$ can be simplified to
$$
dP=\alpha\frac{m_e^2}{p^0}\left(\frac{2\chi}{u}\right)^{1/3}\frac{(u^2+2u+2)\, du\, d\tau}{(1+u)^3}
$$
\begin{equation}\label{SF520}
\times\left[\tilde{y}\, \mbox{Ai}^2(\tilde{y})
+\mbox{Ai}^{\prime\,2}(\tilde{y})\right]+\dots\, ,
\end{equation}
where $\tilde{y}=\left(u/2\chi\right)^{2/3}\tau^2$. Before the averaging over spin state of the initial electron, the expression~(\ref{SF510}) should be completed by adding a term, which depends on the spin-state of the incoming electron~\cite{R70,R79,R86}. However at the limit $\chi\gg 1$, when Eq.~(\ref{SF520}) is valid, this additional spin-dependent contribution becomes much less than the leading term presented in Eq.~(\ref{SF520})~\cite{anom}.

Integrating the latter expression over $\tau$, we obtain the following differential probability for emission of a photon~\cite{R70,R79}:
\begin{equation}\label{SF530}
dP=-\alpha\frac{m_e^2}{p^0}\chi^{2/3}\frac{(u^2+2u+2)\, du}{u^{2/3}(1+u)^3}\mbox{Ai}^{\prime}(0)+\dots\, .
\end{equation}
Integration of the latter expression over $u$ gives the total probability rate of photon emission at $\chi\gg 1$~\cite{R70,R79}:
\begin{equation}\label{SF540}
P=\alpha\frac{m_e^2}{p^0}\chi^{2/3}\frac{14\, \Gamma(2/3)}{3^{7/3}}+\dots
=\frac{\alpha}{p^0}\frac{14\, \Gamma(2/3)}{3^{7/3}}\, f_p^{2/3}+\dots\, ,
\end{equation}
where
\begin{equation}\label{SF010}
f_p^2=
m_e^6\chi^2=e^2\left(F_{\nu\mu}p^\nu\right)\left(F^{\mu\lambda}p_\lambda\right)
\, .
\end{equation}
The dots ``$\dots$'' in Eqs.~(\ref{SF520})-(\ref{SF540}) designate residual terms that do not give contributions to the leading term in Eq.~(\ref{SF540}).

We see that the definition of $f_p$ does not contain electron mass, $m_e$. Thus, particle mass dropped from the expression~(\ref{SF540})~\cite{inv}, if it is considered as function of electron 3-momentum $\pmb{p}$ at $|\pmb{p}|\gg m_e$ and $\chi\gg 1$. This expression for $P$ remains valid for any particles of the spin 1/2 with electric charge $e$, once
\begin{equation}\label{SF550}
m^6 \ll f_p^2
\, ,
\end{equation}
where $m$ is its mass. This conclusion is very important for our further reasoning. It is why we repeated though very shortly main steps of known derivation of the 1st part of Eq.~(\ref{SF540}).

The statement in the previous paragraph is very important and provides a basis for main results of this paper. We give here a sketch of our further consideration based on the conclusion of the previous paragraph.

Let $\tilde{m}_e$ be an exact electron mass, that takes into account all radiation corrections to electron mass in the crossed field. It defines in particular the exact phase of electron wave function in the strong field. The statement of the paragraph, containing Eq.~(\ref{SF550}), can be expressed in the following way. If 
\begin{equation}\label{SF552}
|\tilde{m}_e^6| \ll f_p^2
\, ,
\end{equation}
then the leading order for probability rate of photon emission by electron can be expressed as in Eq.~(\ref{SF540}). Thus Eq.~(\ref{SF540}) takes exactly into account radiation corrections to incoming and outgoing states of electron for the process: $e\to e^\prime+\gamma$. To avoid misunderstanding, we note once again that it concerns only the leading term of Eq.~(\ref{SF540}). We will explain in the next section that this statement remains valid for corrections of the photon mass in the outgoing state of the electromagnetic field, if additionally
\begin{equation}\label{SF554}
|\mu^6| \ll f_p^2
\, ,
\end{equation}
when $\mu^2$ is exact radiation correction to the square of the photon mass in the strong field. We will explain in the next section also that analogous statement is valid for the probability of the process   $\gamma\to e^++e^-$, if 
\begin{equation}\label{SF556}
|\mu^6|\, ,\quad  |\tilde{m}_e^6| \ll f_k^2
\, ,
\end{equation}
where $f_k$ is defined for the photon analogously to (\ref{SF010}), with evident replacement of $p$ by the wave 4-vector of the photon. We will present in the next paragraph arguments in favor, that radiation correction to the vertex for the reaction  $e\to e^\prime+\gamma$ does not destroy the asymptotic behavior of $P$. 

This $P$ will define an imaginary part of $\tilde{m}_e^2$ in this limit and, hence, the whole asymptotics of $\tilde{m}_e^2$. Analogously the probability rate of the process  $\gamma\to e^++e^-$ will define $\mu^2$. It s very important that  $\tilde{m}_e^2$ and  $\mu^2$ determined by this way automatically obey the conditions~(\ref{SF552})-(\ref{SF556}) under condition that 
$f_k^2$ and $f_p^2\gg m_e^6$. This conclusion finalizes logic structure of derivation of main results of this paper. 

The differential probability rate~(\ref{SF510}) of the reaction $e\to e^\prime+\gamma$ is calculated~\cite{R70,R79} via a matrix element of the operator proportional to $j \hat{\cal A}$, where $\hat{\cal A}$ is a vector potential operator of emitted photon, and not the external field. The matrix element is calculated between the Volkov states~\cite{Vol,QED} with momenta $p$ and $p^\prime$ in the external crossed field. The matrix element is determined mainly by integration over space-time domain with the size along the external low frequency wave of the order of the formation length~\cite{R70,R79}: $\ell_c\sim p^0 f_p^{-2/3} (f_{p^\prime}/f_k)^{1/3}$ for $\chi$, $\chi^\prime$ and $\kappa\gg1$. It means that probability of the reaction on the formation length is of the order of $\alpha\, {\cal O}(\chi^\prime/\chi)$~\cite{al}, and it does not contain at all the parameter $\alpha\chi^{2/3}$. The Volkov states do not take into account their radiative decays during the reaction. This neglecting correspond to relative uncertainties in Eq.~(\ref{SF510}) of the order of $\alpha$. Thus the leading order, presented in Eq.~(\ref{SF540}), corresponds to the limit $\alpha\ll 1$, $\chi\gg1$, regardless of any relationship between $\alpha^{-1}$ and $\chi$.

We assume for convenience that $p^0\gg |\tilde{m}_e|$. In this case, a coefficient of damping rate of amplitude of the wave function can be written as
$\Im (-\tilde{m}_e^2)/(2p^0)$, which should be equal to $P/2$ as obtained previously in Eq.~(\ref{SF540}). Here $\Im (z)$ is an imaginary part of the complex number $z$. Hence, we have
\begin{equation}\label{SF560}
\Im(\tilde{m}_e^2) = -p^0 P= -\alpha\, \frac{14\, \Gamma(2/3)}{3^{7/3}}\, f_p^{2/3}+\dots\, .
\end{equation}
Analytic continuation of $\tilde{m}_e^2$ on the whole physical part of the complex space of $f_p$, using the restriction~(\ref{SF560}), gives
\begin{equation}\label{SF570}
\tilde{m}_e^2 = m_e^2+C_e\,\alpha\, f_p^{2/3}+\dots
\quad (\mbox{at~~} f_p^2\gg m_e^6)\, ,
\end{equation}
where  the complex numerical coefficient $C_e$ is defined as
$$
C_e=\frac{28\,\Gamma(2/3)e^{-i\pi/3}}{3^{17/6}}\, .
$$
Typical ways of such analytic continuation of analogous functions in QED were considered in Refs.~\cite{R79,R86}. To be shorter, we assume actually that analytic properties of the expression~(\ref{SF570}) do not depend on the relationship between two additive terms in Eq.~(\ref{SF570}), while $\chi\gg1$. Then we are able to use the results of the papers~\cite{R70,R72} at $\chi\gg1$. We may emphasize once again that the second term in the right hand side of Eq.~(\ref{SF560}) does not depend at all on the unknown yet effective mass of the electron under condition~(\ref{SF552}). We mean here a leading order of $\tilde{m}_e^2$ at $ f_p^2 \gg|\tilde{m}_e|^6$.

The condition of applicability of Eq.~(\ref{SF570}), $f_p^2\gg m_e^6$, is necessary, but in accordance to our previous consideration of the result~(\ref{SF540}), we should demand more restrictive on the 1st glance condition~(\ref{SF552}) for Eq.~(\ref{SF540}) would be valid. However, in view of Eq.~(\ref{SF570}), the strong inequality~(\ref{SF552}) is valid automatically if $f_p^2\gg m_e^6$, because of entering the small parameter $\alpha$ into Eq.~(\ref{SF570}). It is a crucial point of the present work.

Thus, modified effective electron mass in the strong field can be presented as
\begin{equation}\label{SF030}
\tilde{m}_e=\sqrt{m_e^2+C_e\,\alpha\,
f_p^{2/3}}+\dots\, .
\end{equation}
The dots indicate here subleading terms at $f_p^2\gg m_e^6$. The leading term in $\tilde{m}_e-m_e$ determines asymptotics of diagonal, spin-independent term of electron mass operator.
We noted above that  spin-dependent part of $P$ is much less than the leading `scalar' term at $f_p^2\gg m_e^6$. Hence, we may set the same for the spin-dependent part of the mass operator, as well. It is much less than the leading order term, at least close to the mass shell.

We saw that Eq.~(\ref{SF530}) is valid under two separate conditions $\alpha\ll1$ and $f_p^2\gg m_e^6$, regardless of any relationship between
$\chi^2=f_p^2/m_e^6$ and $\alpha^{-1}$. Hence, we say the same about Eq.~(\ref{SF030}). In particular, it remains valid at $\alpha^3f_p^2\gg m_e^6$, when the 2nd term under the square root is much larger than the 1st one: \begin{equation}\label{SF040}
\tilde{m}_e\bigr|_{\alpha^3f_p^2\gg m_e^6}=\sqrt{C_e}\,\sqrt{\alpha}\,
f_p^{1/3}+\dots\, .
\end{equation}
In the opposite case, when $\alpha^3f_p^2\ll m_e^6$ (and, hence, $\alpha\chi^{2/3}\ll1$), but still $f_p^2\gg m_e^6$ (or $\chi\gg1$), the expression~(\ref{SF030}) can be expanded in a Taylor series. Thus, this case can be investigated with a perturbation theory.  Note, that perturbation methods in QED~\cite{QED} are developed for the mass operator and not for the effective mass square. The Taylor series becomes divergent at $\alpha\chi^{2/3}\gtrsim 1$. It is why the result~(\ref{SF030}) cannot be derived straightforwardly by a regular perturbation method.

\section{Invariant photon mass in a strong field at $\kappa\gg1$}

Turning now to the effective mass of photons, we are able to be very short, because our derivation will reproduce almost completely the derivation of the previous section. We start from probability rate of the reaction $\gamma\to e^++e^-$ in the strong crossed field. It was calculated in Refs.~\cite{NR64,NR67,R79}, and it can be presented at $\kappa\gg 1$ in the way that is quite similar to the set of equations~(\ref{SF510})-(\ref{SF540}). Again the final result will be quite similar to Eq.~(\ref{SF540}) with the obvious substitution $f_p\to f_k$ and with other numerical coefficient. The most important feature of this expression is that it does not contain again $m_e$. Hence, it remains valid for the process of production any particle-antiparticle pair, when the particle has the charge $e$ and spin 1/2 under condition that the mass $m$ of the particle fulfills the following strong inequality
\begin{equation}\label{SF060}
m^6 \ll f_k^2\, .
\end{equation}
In particular, considering  $e^+e^-$-pair production in the strong electromagnetic field, we assume
\begin{equation}\label{SF060}
|\tilde{m}_e^6| \ll f_k^2\, .
\end{equation}

Hence, we may conclude that it remains valid for $e^-e^+$-pair production, when we take into account modification of the electron mass in the same electromagnetic field. We may again neglect production of $e^-$ or $e^+$ with $f_p\sim |\tilde{m}_e|^3$ or with parameter $f_{p^\prime}\sim |\tilde{m}_e|^3$, respectively, because such reactions will give only a subleading contribution to the probability rate. As a result, we may conclude that the known result~\cite{R70,N69,R79} for the invariant mass of photons in the strong field
remains valid regardless of any relationships between $\kappa$ and $\alpha^{-1}$. Thus,
\begin{equation}\label{SF070}
\mu_{\parallel,\, \perp}^2\bigr|_{f_k^2\gg m_e^6}=C_{\gamma\, \parallel,\, \perp}\, \alpha f_k^{2/3} \, ,
\end{equation}
where
\begin{equation}\label{SF080}
C_{\gamma\, \parallel,\, \perp}=\frac{(5\mp1)\, 3^{7/6}}{14\pi^2}\, \Gamma^4(2/3)\, e^{-i\pi/3}\, .
\end{equation}
Here $\parallel$ and $\perp$ designate polarization of the incoming $\gamma$-photon in a special reference frame, where external high intensity low frequency wave and the photon are counter-propagate to each other.

We note that, though the expression in the right hand side of Eq.~(\ref{SF070}) coincides with the result~\cite{R70} of the perturbation theory, it remains valid in the region, where the perturbation theory does not work, especially because we use the result~(\ref{SF030}) that,  rigorously speaking, cannot be obtained in the the perturbation theory.

There is an attractive question. What happens, if we take into account radiative correction~(\ref{SF070}) of the photon mass in the calculation of probability rate~(\ref{SF510}) of the reaction $e\to\gamma+e^\prime$? We may check, that this procedure will result mainly in appearing a contribution
$\sim \mu_{\parallel,\, \perp}^2\chi^2/(m_e^2 \kappa^2) \sim \mu_{\parallel,\, \perp}^2/m_e^2$
to the term $1+\tau^2$ in the definition~(\ref{SF512}) of $y$. 
An origin of the term $\sim \mu_{\parallel,\, \perp}^2/m_e^2$ is quite similar to the origin of the term 1 in the expression $1+\tau^2$. The latter one comes from the phase rate imbalance $i(k^x-p^x+p^{\prime x})(x+t)$ in potentially highly oscillating exponent, when we calculate the matrix element of the interaction $e\to e^\prime+\gamma$ . The imbalance is  caused by finite electron mass. We use here a special frame of reference where emitting electron exactly counter-propagates the strong wave (that is actually constant crossed field). These contributions appear generally even when 3-momenta of photon and electron after emission are parallel to the initial 3-momentum of the electron. Applying conservation laws, the phase rate imbalance becomes equal to $i(p^{\prime 0}+k^{\prime 0}-p^{0})(x+t)$. Combining these imbalance phase rate with potentially highly oscillating phase rate contribution,  caused by influence of the strong field on the electrons that is proportional to $(x+t)^3$, and taking in mind mass of electrons and photons, we obtain the term 1 in the definition of $y$ and the term $\sim \mu_{\parallel,\, \perp}^2/m_e^2$, caused by finite photon mass.
The correction $\sim \mu_{\parallel,\, \perp}^2/m_e^2 $
gives contribution $\sim\alpha$ to $y$ and will get lost at the transition from Eq.~(\ref{SF510}) to Eq.~(\ref{SF520}) and then to Eq.~(\ref{SF540}). The same conclusion can be drawn for the process $\gamma\to e^++e^-$ at $\kappa\gg1$, if we take into account finite $\mu_{\parallel,\, \perp}^2$ for the incoming photon.

The arguments of the previous paragraph means that the leading order corrections to propagators of electrons and photons, described by corrections~(\ref{SF030}) and~(\ref{SF070}) to their masses are self consistent at the limit $\kappa$, $\chi\gg1$. It means in particular that they are self consistent for arbitrary large $\alpha\kappa^{2/3}$ and $\alpha\chi^{2/3}$.

\section{Discussion and Conclusions}

We derived Eqs.~(\ref{SF030}) and~(\ref{SF070}) for modification of effective electron and photon masses in a strong crossed electromagnetic field. We argued that they remains valid for arbitrary high $\chi\gg1$ and $\kappa\gg1$. Our arguments come from the observation that invariant probability rates of the processes $e\to e^\prime+\gamma$ and
$\gamma\to e^++e^-$ in their leading orders does not depend at all on the masses, if $\kappa\gg1$ and $\chi\gg1$, and the masses are determined by Eqs.~(\ref{SF030}) and~(\ref{SF070}). The leading orders are determined by a range of parameters, where  $f_k^2$ and $f_p^2$ are much larger than $|\tilde{m}_e^6|$ and $\mu^2$ for outgoing particles also. However, if we would be interesting especially for example   the process $e\to e^\prime+\gamma^\prime$, when $\bigl|f_{k^\prime}^\prime\bigr| \lesssim |\tilde{m}_e^3|$, that represents in particular the SCCR process~\cite{B19}, then the partial probability of this process may be strongly modified at $\alpha \chi^{2/3}\gtrsim 1$ and unlikely could be considered with a standard perturbation theory.

These results are based on the expression~(\ref{SF540}) for asymptotics of the probability rate, $P$, of the process $e\to e^\prime+\gamma$ in the strong field and on the analogous asymptotics of probability rate of the process $\gamma\to e^-+e^+$. We argued that that they remain valid as asymptotics when $f_p^2\gg m_e^6$ and $f_k^2\gg m_e^6$ regardless of values of the ratios  $|\tilde{m}_e-m_e|/m_e$ and $\mu^2/m_e^2$. We did not use any perturbation theories based on smallness of these ratios, using standard QED perturbation theory based on smallness of the fine structure constant, $\alpha$. We may say that the Furry picture was only a starting point for evaluation of asymptotics of the probabilities of the processes mentioned above, whereas our final interpretation of the expression is not based on the Furry picture.

Our results may help to resolve very long-lasted controversy of the perturbation theory for QED in a strong external electromagnetic field. It was reviewed briefly in our Introduction. The result cannot be obtained straightforwardly from this perturbation theory. We may hope that it could be obtained, maybe more rigorously, by the following way: i) by using of the perturbation theory at small $\alpha\chi^{2/3}$ and $\alpha\kappa^{2/3}$, and ii) by means of summation of infinite number of diagrams, giving most contribution to the divergence at finite $\alpha\chi^{2/3}$ and $\alpha\kappa^{2/3}$, and iii) by consequent analytic continuation of the result of summation to higher $\alpha\chi^{2/3}$ and $\alpha\kappa^{2/3}$. See for example~\cite{Mi20,Mi21} and references therein. We believe that our derivation of Eqs.~(\ref{SF030}) and~(\ref{SF070}) may help to insight into this problem.  

Our main result can be expressed also as follows. Expressions for electron and photon mass squares, modified due to  creation of virtual electron-positron pairs and photons in strong crossed electromagnetic field, and obtained in the frame with regular perturbation theory (in strong external field) in the limit $1\ll \chi$ and $\kappa\ll \alpha ^{-3/2}$ remain actually valid in its leading order for arbitrary large $\chi$ and $\kappa$, where the perturbation theory breaks actually.

Main `practical' consequence of our results is that the expressions~(\ref{SF520})-(\ref{SF540}) for the rate of reaction $e\to e^\prime+\gamma^\prime$ in a strong electromagnetic field that are valid at $\chi\gg 1$ remain valid at arbitrary high values of $\alpha\chi^{2/3}$ if we neglect rare events of low energies of the secondary particles, $e^\prime$ and $\gamma^\prime$. This process is actually multi-photon Compton scattering with complete accounting for radiation corrections. This consequence was used for simulation of highly energetic electron achieving of extremely strong electromagnetic field generated by optical lasers and of radiation of such electrons~\cite{Ji22}. Experimental confirmation of the predictive simulations of Ref.~\cite{Ji22} may confirm absence of a physical peculiarities at $\alpha\chi^{2/3}$ and $\alpha\kappa^{2/3}\sim 1$.  A quite analogous conclusion can be drawn for the rate of pair production by a photon ($\gamma\to e^-+e^+$) with $\kappa\gg1$ in a strong electromagnetic field. This process is actually the multi-photon Breit-Wheeler process.

Present experimental techniques move gradually but permanently  to achieving the region, where $\alpha\chi^{2/3}$ and $\alpha\kappa^{2/3}$ will become higher than  $~1$. Modern standard accelerators and high intensity lasers may be used to reach this region~\cite{F18,Fed21,Ji22}. Then our results may be checked experimentally.

\begin{acknowledgments}

We acknowledge our financial support from the project High Field Initiative (CZ.02.1.01/0.0/0.0/15\_003/0000449)
from the European Regional Development Fund.
\end{acknowledgments}


\begin{thebibliography}{99}
\bibitem{un} We use throughout the whole paper a Gaussian system of units with $\hbar=c=1$.
\bibitem{inv} Meanwhile, already Eq.~(\ref{SF520}) can be rewritten in the form that demonstrates evidently that total invariant probability does will not contain $m_e$:
$$
dP=\frac{\alpha}{p^0}(2f_p)^{2/3}\left\{\left[\psi^2\, \mbox{Ai}^2(\psi^2)
+\mbox{Ai}^{\prime\,2}(\psi^2)\right]\, d\psi\right\}
$$
$$
\times\frac{(u^2+2u+2)\, du}{u^{2/3}(1+u)^3}\, ,
$$
where $\psi=(u/2\chi)^{1/3}\tau$.
\bibitem{al} To be more rigorous, we may express this statement in the following way: $\ell_c(\chi,\chi^\prime)\,
u\, [dP(\chi,\chi^\prime)/du]=\alpha\, {\cal O} (\chi^\prime/\chi)$.
\end{thebibliography}
\end{document}